\documentclass{article}
\usepackage{locata,amsmath,graphicx,url,times}
\usepackage{booktabs}
\usepackage{balance}

\title{LOCALIZATION AND TRACKING OF AN ACOUSTIC SOURCE USING A DIAGONAL UNLOADING BEAMFORMING AND A KALMAN FILTER}

\name{Daniele Salvati, Carlo Drioli, Gian Luca Foresti}

\address{Department of Mathematics, Computer Science and Physics \\
			University of Udine\\
       via delle Scienze, 206, 33100 Udine, Italy}

\begin{document}

\ninept
\maketitle

\begin{sloppy}

\begin{abstract}
We present the signal processing framework and some results for the IEEE AASP challenge on acoustic source localization and tracking (LOCATA). The system is designed for the direction of arrival (DOA) estimation in single-source scenarios. The proposed framework consists of four main building blocks: pre-processing, voice activity detection (VAD), localization, tracking. The signal pre-processing pipeline includes the short-time Fourier transform (STFT) of the multichannel input captured by the array and the cross power spectral density (CPSD) matrices estimation. The VAD is calculated with a trace-based threshold of the CPSD matrices. The localization is then computed using our recently proposed diagonal unloading (DU) beamforming, which has low-complexity and high resolution. The DOA estimation is finally smoothed with a Kalman filer (KF). Experimental results on the LOCATA development dataset are reported in terms of the root mean square error (RMSE) for a 7-microphone linear array, the 12-microphone pseudo-spherical array  integrated  in  a  prototype  head  for  a  humanoid robot, and the 32-microphone  spherical array.
\end{abstract}


\section{Introduction}
\label{sec:intro}

The aim of an acoustic source localization and tracking system is to estimate the position of sound sources in space by analyzing the
sound field with a microphone array, a set of microphones arranged to capture the spatial information
of sound. Speaker spatial localization/tracking using microphone arrays is of
considerable interest in applications of teleconferencing systems,
hands-free acquisition, human-machine interaction, recognition, and audio surveillance.

In this paper, we present the signal processing framework for the IEEE AASP challenge on acoustic source localization and tracking (LOCATA) \cite{LOCATA2018}. We also present some performance results related to the LOCATA development dataset. 
The proposed localization and tracking system is designed for the direction of arrival (DOA) estimation in single-source scenarios. The localization algorithm is based on diagonal unloading (DU) beamforming, recently introduced in \cite{Salvati2018}. Broadband DU localization beamformer is computed in the frequency-domain \cite{Benesty2007a} by calculating the steered response power (SRP)  on each frequency bin and by summing the narrowband components with the incoherent frequency fusion \cite{Salvati2014}. The tracking is performed with a Kalman filter (KF) \cite{Kalman1960}.

\section{Method}

The proposed system consists of four main building blocks: 
\begin{itemize}
\item pre-processing;
\item voice activity detection (VAD);
\item localization; 
\item tracking.
\end{itemize}
The organization of the signal processing components is illustrated in Figure \ref{system}. 

\subsection{Pre-Processing}

The signal pre-processing pipeline includes the short-time Fourier transform (STFT) of the multichannel input captured by the array $x_m(t)$ ($m=1,2,\dots,M$, where $M$ is the number of microphones). It can be expressed as
\begin{equation}
X_m(k,f)=\sum_{l=-\frac{L}{2}}^{l=\frac{L}{2}-1} w(l)x_m(l+k R)e^{\frac{-j2\pi fl}{L}}, \quad k=0,1,\dots,
\end{equation}
where $k$ is the frame time index, $f$ is the frequency bin, $w(l)$ is the analysis window, $L$ is the size of the fast Fourier transform (FFT), and $R$  is the hop size.

After the frequency-domain transformation, the cross power spectral density (CPSD) matrices  $\mathbf{\Phi}(k,f)$ of the considered frequency range [$f_\text{min}$,$f_\text{max}$] are estimated through the averaging of the array signal blocks \cite{Zhang2011}
\begin{equation}
\begin{split}
&\widehat{\mathbf{\Phi}} (k,f)=\frac{1}{N}\sum_{k_n=0}^{N-1} \mathbf{x}(k-k_n,f)\mathbf{x}^H(k-k_n,f), \\
&f=f_\text{min},f_\text{min}+1,\dots, f_\text{max},
\end{split}
\label{ecm}
\end{equation}
where $N$ is the number of frames for the averaging, $H$ denotes the conjugate transpose operator, and 
\begin{equation}
\mathbf{x}(k,f)=[X_1(k,f), X_2(k,f), \dots, X_M(k,f)]^T,
\end{equation}
where $T$ denotes the transpose operator.

\begin{figure*}[t]
\centerline{\includegraphics[width=2.0\columnwidth]{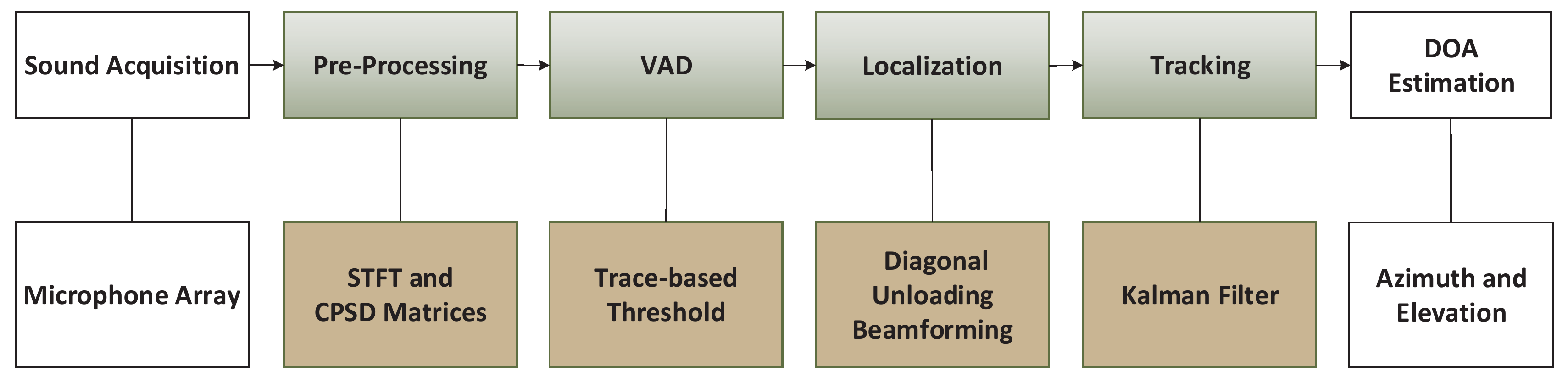}}
\caption{Schematic diagram of the proposed system.}
\label{system}
\end{figure*}

\subsection{VAD}

The VAD used herein is based on the trace of the CPSD matrices that is related on the DU beamforming. The trace of a CPSD matrix is equivalent to the sum of the eigenvalues of the matrix, i.e., it represents the overall power of the array. The source detection is hence calculated as
\begin{equation}
  \text{VAD}(k)=
  \begin{cases}
    1, & \text{if $\sum_{f=f_\text{min}}^{f_\text{max}} \text{tr}[\widehat{\mathbf{\Phi}} (k,f)]>\eta$},\\
    0, & \text{otherwise},\\
  \end{cases}
	\label{rms}
\end{equation} 
where $\text{tr}[\cdot]$ is the operator that computes the trace of a matrix, and $\eta$ is a given threshold. The parameter $\eta$ was empirically
set to the value allowing to effectively detect the source activity.   

\subsection{Localization}

The acoustic source DOA estimation method is a low complexity and robust beamformer based on a DU transformation of the covariance matrix involved in the conventional beamformer computation to exploit the high resolution subspace orthogonality property. The method is illustrated in details in \cite{Salvati2018}.

The transformation, on which the DU method is based, is obtained by subtracting an opportune diagonal matrix from the CPSD matrix $\widehat{\mathbf{\Phi}} (k,f)$ of the array output vector. As a result, the DU beamforming removes as much as possible the signal subspace from the covariance matrix and provides a high resolution beampattern. In practice, the design and implementation of the DU transformation is simple and effective, and is obtained by computing the matrix (un)loading factor.

The broadband SRP is defined as \cite{Salvati2018,Salvati2014}
\begin{equation}
P(k,\mathbf{\Omega}_d)=\sum_{f=f_\text{min}}^{f_\text{max}} \frac{P_\text{DU}(k,f,\mathbf{\Omega}_d)}{||\mathbf{g}(k,f)||_{\infty}},
\label{pfn}
\end{equation}
where $\mathbf{\Omega}_d=[\theta_d,\phi_d]$  ($\theta_d$ and $\phi_d$ are the azimuth and elevation angles) is the steering direction, $||\cdot||_{\infty}$ denotes the Uniform norm, i.e., the maximum value of the vector 
\begin{equation}
\mathbf{g}(k,f)=[P_\text{DU}(k,f,\mathbf{\Omega}_1),P_\text{DU}(k,f,\mathbf{\Omega}_2),\dots,P_\text{DU}(k,f,\mathbf{\Omega}_D)],
\end{equation}
which contains all the narrowband SRP for the considered search direction $D$, and the narrowband DU response power beamforming $P_\text{DU}(k,f,\mathbf{\Omega}_d)$ is defined as
\begin{equation}
P_\text{DU}(k,f,\mathbf{\Omega}_d)=\frac{1}{\mathbf{a}^H(f,\mathbf{\Omega}_d) [\text{tr}[\widehat{\mathbf{\Phi}} (k,f)] \mathbf{I}-\widehat{\mathbf{\Phi}} (k,f)] \mathbf{a}(f,\mathbf{\Omega}_d)},
\label{du}
\end{equation}
where $\mathbf{a}(f, \mathbf{\Omega}_d)$ is the array steering vector for the direction  $\mathbf{\Omega}_d$, and $\mathbf{I}$ is the identity matrix. Note that the unloading parameter is computed with the trace operation of the CPSD matrices. This solution guarantees that the transformed PSD matrix $\mathbf{\Phi}_\text{DU}(k,f)=[\text{tr}[\widehat{\mathbf{\Phi}} (k,f)] \mathbf{I}-\widehat{\mathbf{\Phi}} (k,f)]$ has the attenuation of the signal subspaces with respect to the noise subspace, and hence the high resolution orthogonality is exploiting, even if partially, since the transformed PSD matrix is affected by a
certain amount of signal subspace \cite{Salvati2018}. The array steering vector depends on the array geometry. Note that for the linear array the steering direction is given only by the azimuth angle.

Then, the DOA estimate of the source is obtained by 
\begin{equation}
\hat{\mathbf{\Omega}}_s(k)=\underset{\mathbf{\Omega}_d}{\operatorname{argmax}} [P(k,\mathbf{\Omega}_d)], \quad d=1,2,\dots,D.
\end{equation}

\subsection{Tracking}

The KF \cite{Kalman1960} is an optimal recursive Bayesian filter for
linear systems observed in the presence of Gaussian noise. The filter equations can be divided into a prediction and a correction
step. The state of the process is given by
\begin{equation}
\mathbf{y}(k)=[\mathbf{\Omega}(k), v_{\theta}(k), v_{\phi}(k)]^T,
\end{equation}
where $v_{\theta}(k)$ and $v_{\phi}(k)$ are the velocities.
In the prediction step the update equations are
\begin{equation}
\mathbf{y}_p(k)=\mathbf{A} \mathbf{y}(k-1),
\end{equation}
\begin{equation}
\mathbf{P}_p(k)=\mathbf{A}\mathbf{P}(k-1)\mathbf{A}^T+\mathbf{B}\mathbf{Q}\mathbf{B}^T,
\end{equation}
where
\begin{equation}
\mathbf{A}=
\begin{bmatrix}
1 & 0 & dt & 0 \\
0 & 1 & 0 & dt \\
0 & 0 & 1 & 0 \\
0 & 0 & 0 & 1 \\
\end{bmatrix},
\end{equation}
\begin{equation}
\mathbf{B}=
\begin{bmatrix}
0.5dt^2 & 0  \\
0 & 0.5dt^2\\
dt & 0\\
0 & dt\\
\end{bmatrix},
\end{equation}
\begin{equation}
\mathbf{Q}=
\begin{bmatrix}
\sigma_q^2 & 0  \\
0 & \sigma_q^2 \\
\end{bmatrix},
\end{equation}
with $\sigma_q^2$ being the variance of the process error, $dt=RN/f_s$ the time  elapsed between DOA estimations, $f_s$ the sampling rate. The filter is initialized with the state covariance matrix $\mathbf{P}(k_i)=\mathbf{B}\mathbf{Q}\mathbf{B}^T$ and the state $\mathbf{y}(k_i)=[\hat{\mathbf{\Omega}}_s(k_i),0, 0]^T$, where $k_i$ is the first time frame in which the VAD($k_i$) has value 1 and VAD($k_i$-1)=0. 
After the prediction step, the Kalman gain is calculated as
\begin{equation}
\mathbf{K}=\mathbf{P}_p(k) \mathbf{C}^T(\mathbf{C} \mathbf{P}_p(k) \mathbf{C}^T + \mathbf{R})^{-1},
\end{equation}
where
\begin{equation}
\mathbf{C}=
\begin{bmatrix}
1 & 0 & 0 & 0 \\
0 & 1 & 0 & 0\\
\end{bmatrix},
\end{equation}
\begin{equation}
\mathbf{R}=
\begin{bmatrix}
\sigma_r^2 & 0  \\
0 & \sigma_r^2 \\
\end{bmatrix},
\end{equation}
with $\sigma_r^2$ being the variance of the measurement error.
In the correction step the measurement update equations are
\begin{equation}
\mathbf{y}(k)=\mathbf{y}_p(k)+\mathbf{K}(\hat{\mathbf{\Omega}}_s(k)-\mathbf{C} \mathbf{y}_p(k)),
\end{equation}
\begin{equation}
\mathbf{P}(k)=(\mathbf{I} - \mathbf{K}\mathbf{C})\mathbf{P}_p(k).
\end{equation}
Hence, after the correction step the filtered DOA estimation $\hat{\mathbf{\Omega}}^\text{EKF}_s(k)=\mathbf{\Omega}(k)$ is obtained.

\section{Experimental Results}

We present some experimental results on the LOCATA development dataset to show the performance of the proposed framework in the single-source scenario with:
\begin{itemize}
\item static loudspeaker and static array (task 1);
\item moving speaker and static array (task 3); 
\item moving speaker and moving array (task 5). 
\end{itemize}
We tested the system with the distant talking interfaces for control of interactive TV (DICIT) array by considering a 7-microphone linear subarray ([4 5 6 7 9 10 11]) taking into account the far-field model, the 12-microphone pseudo-spherical array integrated  in  a  prototype  head  for  a  humanoid robot array, and the 32-microphone eigenmike spherical array. The system setup is implemented with the following parameters: 
\begin{itemize}
\item sampling rate: 48 kHz;
\item STFT window: Hann function $w(l)$;
\item FFT size: $L=2048$ samples;
\item hop size: $R=512$ samples;
\item number of frames for CPSD estimation: $N=25$;
\item frequency range: [$f_\text{min}$,$f_\text{max}$]=[80,8000] Hz;
\item VAD threshold: $\eta=200$ (linear array), $\eta=50$ (robot head), $\eta=10$ (eigenmike);
\item spatial resolution: 1 degree (linear array, $D=181$), 5 degrees (robot head and eigenmike, $D=2701$);
\item DOA estimation time period: $dt=0.2667$ s;
\item KF parameters: $\sigma_q^2=10^{-3}$, $\sigma_r^2=10^{-4}$. 
\end{itemize}
\balance 
The signal processing framework has been implemented using Matlab R2017a. We used our own implementation for the KF.
The performance was assessed in terms of the root mean square error (RMSE). Table \ref{res} shows the DOA estimation results for each task and each recording. The azimuth angle was evaluated for the linear array, while both azimuth and elevation angles was considered for the robot head and eigenmike array. Three examples of detection, localization and tracking are depicted in Figures \ref{alg_dicit_task1_recording3}, \ref{alg_benchmark2_task3_recording2}, \ref{alg_eigenmike_task5_recording1}. Figure \ref{alg_dicit_task1_recording3} shows the performance of the linear array for the task 1 (static loudspeaker, static array) and recording 3. Figure \ref{alg_benchmark2_task3_recording2} shows the performance of the robot head array for the task 3 (moving speaker, static array) and recording 2. Figure \ref{alg_eigenmike_task5_recording1} shows the performance of the eigenmike array for the task 5 (moving speaker, moving array) and recording 1. The top plot shows the waveform of channel 1 with the speaker activity (red line).

\begin{table*}[t]
\renewcommand{\arraystretch}{1.0}
\centering
{\normalsize
\caption{The RMSE (degree) of the localization performance on the LOCATA development dataset.}
\label{res}
\centering
\begin{tabular}{@{}cc|c|cc|cc@{}}
\toprule
& & \multicolumn{1}{c|}{\textbf{Linear array}} & \multicolumn{2}{c|}{\textbf{Robot head}} & \multicolumn{2}{c}{\textbf{Eigenmike}}\\
\midrule
 & & \text{Azimuth} & \text{Azimuth}  & \text{Elevation} & \text{Azimuth}  & \text{Elevation}\\
\midrule
\text{task 1}	& \text{recording 1}	& 0.972		& 1.649	& 2.447	& 5.863	& 2.444\\
	& \text{recording 2}	&5.096	&	0.038	&1.013&	6.676&	6.054\\
	& \text{recording 3}&	1.437		&2.998&	1.980	&7.491&	5.203\\
	\midrule
\text{task 3} &	\text{recording 1}&	6.480	&	3.596&	2.326	&9.939&	3.232\\
	& \text{recording 2}	&9.638	&	4.583&	3.798	&14.244	&4.348\\
	& \text{recording 3}&	4.355	&	2.880&	2.807&	9.370&	5.804\\
	\midrule
\text{task 5}	& \text{recording 1}&	4.912	&	2.338&	1.818	&4.433	&3.100\\
	& \text{recording 2}&	21.196	&	30.217	&11.333&	32.942&	5.738\\
	& \text{recording 3}&	3.086	&	23.010	&7.782	&10.203&	3.473\\
\bottomrule
\end{tabular}}
\end{table*}

\begin{figure*}[t]
\centerline{\includegraphics[width=2.0\columnwidth]{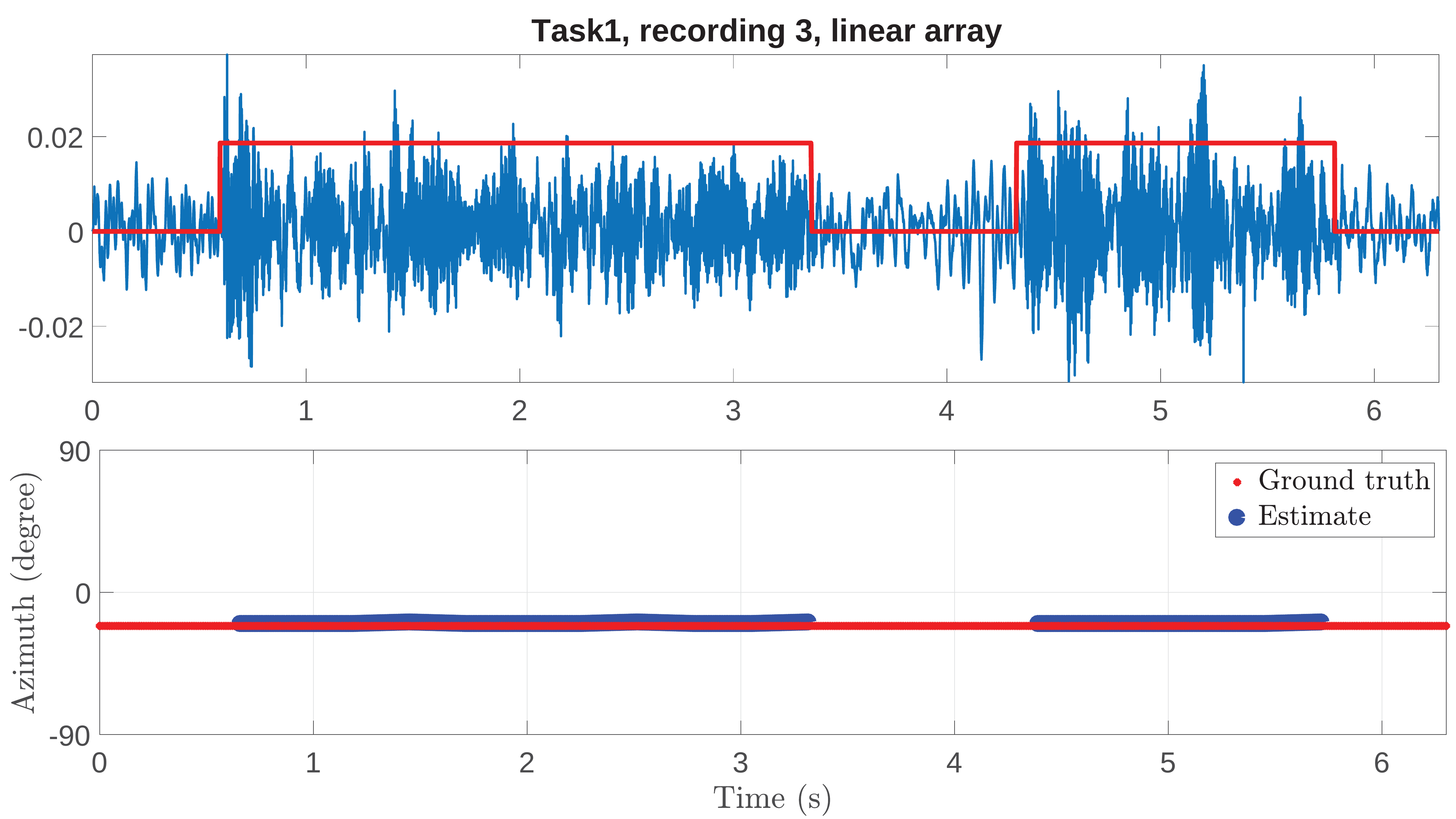}}
\caption{The performance of the proposed system with the 7-microphone DICIT linear subarray for task 1 (static loudspeaker, static microphone array, recording 3).}
\label{alg_dicit_task1_recording3}
\end{figure*}

\begin{figure*}[t]
\centerline{\includegraphics[width=2.0\columnwidth]{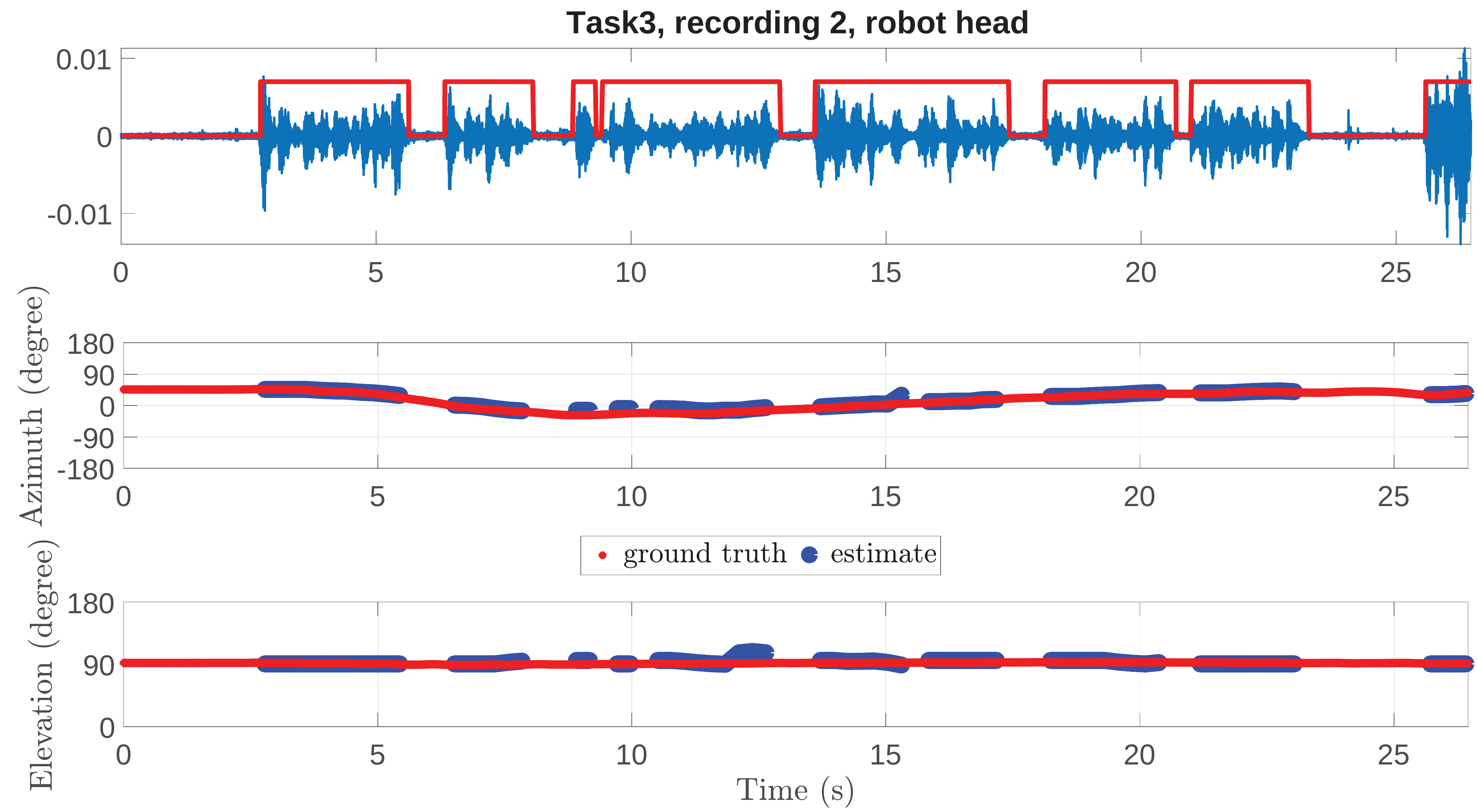}}
\caption{The performance of the proposed system with the robot head array for task 3 (moving speaker, static microphone array, recording 2).}
\label{alg_benchmark2_task3_recording2}
\end{figure*}

\begin{figure*}[t]
\centerline{\includegraphics[width=2.0\columnwidth]{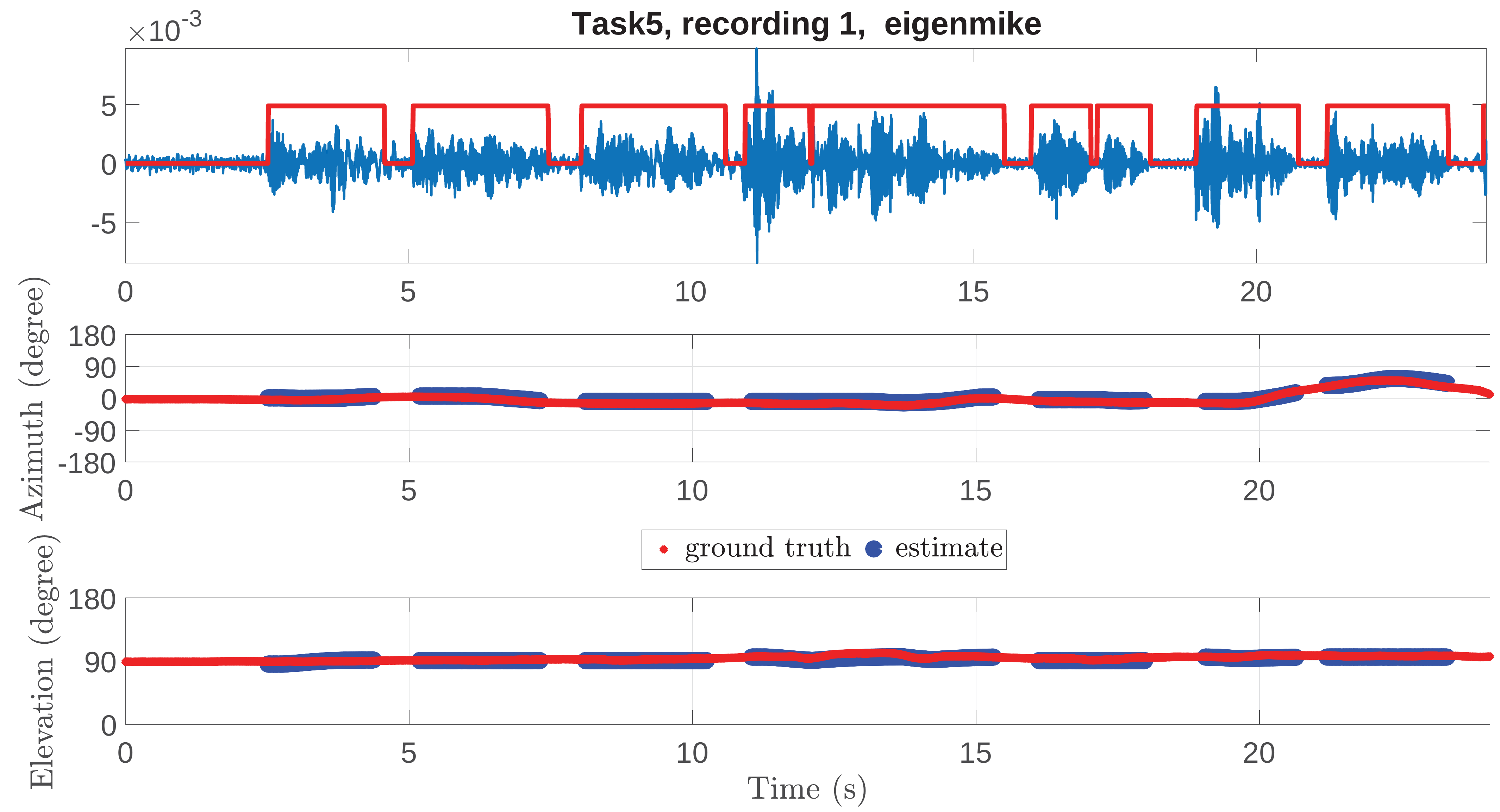}}
\caption{The performance of the proposed system with the eigenmike array for task 5 (moving speaker, moving microphone array, recording 1).}
\label{alg_eigenmike_task5_recording1}
\end{figure*}

\section{Conclusions}

The signal processing framework based on a DU beamforming and a KF for the IEEE AASP LOCATA challenge has been presented. We described the four main building blocks (pre-processing, VAD, localization, tracking) for the DOA estimation of a single source. We showed some results with the LOCATA development dataset using a linear array, the robot head pseudo-spherical array, and the eigenmike spherical array.
\bibliographystyle{IEEEtran}
\bibliography{aslt_du}

\end{sloppy}
\end{document}